\documentclass[a4paper]{jpconf}

\def\apj{ApJ}%
\def\apjl{ApJ}%
\def\apjs{ApJS}%
\def\aap{A\&A}%
\def\ssr{Space~Sci.~Rev.}%
\usepackage{graphicx}
\begin{document}
\title{Deeply X-Raying the High-Energy Sky}

\author{Eugenio Bottacini}
\address{W. W. Hansen Experimental Physics Laboratory and Kavli Institute for Particle Astrophysics and Cosmology,
        Stanford University, USA}
\ead{eugenio.bottacini@stanford.edu}

\author{Marco Ajello}
\address{Department of Physics and Astronomy, Clemson University, Clemson, SC 29634-0978, USA}

\begin{abstract}
All-sky explorations by {\em Fermi}-LAT have revolutionized our view of the gamma-ray Universe. While its ongoing all-sky survey counts thousands of sources, essential issues related to the nature of unassociated sources call for more sensitive all-sky surveys at hard X-ray energies that allow for their identification. This latter energy band encodes the hard-tail of the thermal emission and the soft-tail of non-thermal emission thereby bridging the non-thermal and thermal emission mechanisms of gamma-ray sources.
All-sky surveys at hard X-rays are best performed by current coded-mask telescopes {\em Swift}/BAT and {\em INTEGRAL}/IBIS. To boost the hard X-ray all-sky sensitivity, we have developed an ad hoc technique by combining photons from independent observations of BAT and IBIS. The resulting {\em Swift--INTEGRAL} X-ray (SIX) survey has an improved source-number density. This improvement is essential to enhance the positive hard X-ray -- gamma-ray source matches. We present the results from the scientific link between the neighboring gamma-ray and hard X-ray bands in the context of galactic and extragalactic source classes of the second catalog {\em Fermi} Gamma-ray LAT (2FGL).
\end{abstract}

\section{Introduction}
Whenever a frequency range of the electromagnetic spectrum is newly accessed by  observatories,
the discovery space is huge. So it is for the Large Area Telescope (LAT) \cite{atwood09} of NASA's {\em Fermi}
mission. However, when it comes to astronomical source population studies, data from the single new observatory are in need of a multifrequency approach as the new detected sources must be identified and characterized.
The {\em Fermi}-LAT sources are revealed by the detection of their non-thermal photons. The association with sources at lower frequencies, where emission is mostly thermal, is therefore a major hurdle. Here we show
the association of the {\em Fermi}-LAT sources from the Fermi Second Source Catalog (2FGL) \cite{nolan12}
with those detected in hard X-ray surveys by the {\em INTEGRAL} mission \cite{winkler03} and
by the {\em Swift} mission \cite{gehrels04}. These latter surveys detect the soft tail of the non-thermal photons and the hard-tail of thermal photons form the sources at energies above 15 keV. Thereby they serve as a bridge between non-thermal and thermal emission processes in sources. This allows characterizing and studying the properties of the associated sources.\\
The 2FGL contains 1873 sources detected and characterized in the 100 MeV to 100 GeV range. In this catalog the authors first associate sources depending primarily on close positional correspondence. Some sources are identified by measurements of correlated variability at other wavelength. The density of unassociated {\em Fermi}-LAT sources sharply rises toward the Galactic plane leaving this sky area exposed to large uncertainties in population studies. Sophisticated association studies of the 2FGL at radio and IR frequencies show promising results for blazars on the extragalactic sky \cite{massaro12,schinzel15}, while the Galactic plane and other types of sources are difficult to access at these frequencies. There are also a number of methodical and very precise follow-up observations at soft X-ray energies ($<$ 10 keV) with {\em Swift}/XRT available \cite{stroh13}. These are all predetermined and identified sources from the Fermi-LAT catalog. Single sources are followed-up also with NuSTAR \cite{tendulkar14}. However, these follow-up observations with focusing instruments (e.g. {\em NuSTAR}, {\em Swift}/XRT, etc.) pinpoint sources that are known in advance and systematic properties of the {\em Fermi}-LAT sources are difficult to derive because of the absence of the characteristics on the whole source population. Also follow-up observations with focusing telescopes are useful only for {\em Fermi}-LAT sources whose positional uncertainty is less than the extent of the field of view of the focusing telescope itself. An explanatory example for this is the {\em Swift}/XRT follow-up observation (id 00032313001) of the LAT source J0902-4624 \cite{ojha12}. We have analyzed the follow-up observation with {\texttt{xrtproducts} and {\em HEAsoft 6.11.1}. In Figure~\ref{fig:lat-xrt} we display the XRT intensity map where the red dashed circle and green solid circle centered in RA=135.56 deg, Dec=-46.41 deg represent a radius of 0.25 deg and 1 deg containment circle corresponding to 68\% and 95\% confidence level (CL) respectively. It is worth noting that the XRT observation does not even cover the 68\% CL, while to cover the more realistic 95\% CL $\sim$20 XRT tiles would be necessary. A further complication comes from the mostly thermal photons from the multiple detected sources by XRT. These photons do not allow for a spectral bridge to the non-thermal photons detected by the LAT.\\

\begin{figure}[h]
\includegraphics[width=18pc]{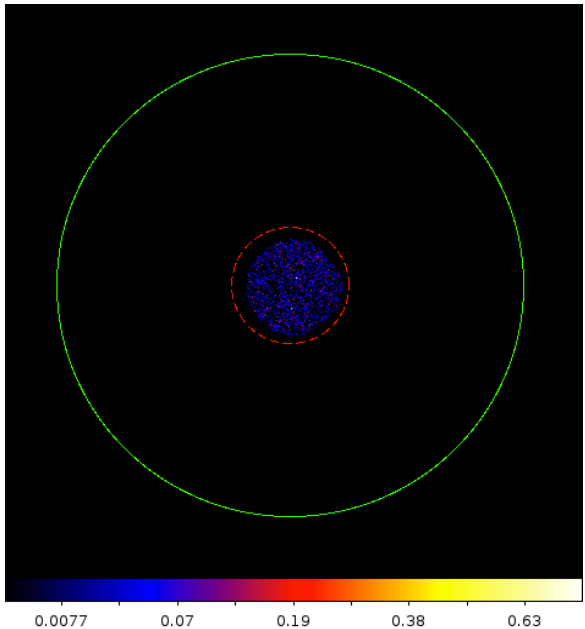}\hspace{2pc}%
\begin{minipage}[b]{14pc}\caption{\label{fig:lat-xrt}XRT observation enclosed in red dashed and green solid circles that represent 68\% and 95\% CL containment radius.}
\end{minipage}
\end{figure}

\section{Current hard X-ray surveys}
To associate {\em Fermi}-LAT sources independent of the sky area (whether extragalactic or Galactic), of frequencies, and of pre-selection method is to use all-sky surveys at hard X-ray energies with coded-mask telescopes. In fact, these telescopes provide surveys over the whole sky encoding non-thermal and thermal emission, and they do not suffer from pre-selection of sources. Many successful studies \cite{beckmann06,sazonov07,tueller08,ajello08,cusumano10,ajello12,krivonos12,bottacini12,baumgartner13} proof their effectiveness of such surveys with coded-mask telescopes. The drawback of the coded-mask imaging systems is that they suffer from a limited sensitivity issue because, by optical design, 50\% of the incident photons are block by the coded-mask itself.\\
To address the limited sensitivity issue of coded-mask telescopes, we have developed a new and ad-hoc technique that consists in summing up the independent photons detected by the Burst Alert Telescope \cite{barthelmy05} of the {\em Swift} mission and the {\em INTEGRAL} Soft Gamma-Ray Imager (ISGRI) \cite{lebrun03} of the Imager on Board the {\em INTEGRAL} Satellite (IBIS) \cite{ubertini03}. The resulting {\em Swift}-{\em INTEGRAL} X-ray (SIX) survey is more sensitive than the surveys of the two instruments alone. A full description of this technique can be found in \cite{bottacini12,bottacini13}. The improved sensitivity makes the SIX survey very suited to detect weak sources and find the counterparts of gamma-ray sources.\\

\section{Finding the hard X-ray counterparts of the {\em Fermi}-LAT sources}
The first attempt to systematically associate {\em Fermi}-LAT sources to counterparts in the hard X-ray energy domain was performed in a detailed study by \cite{ubertini09} using the {\em INTEGRAL}/IBIS survey alone. The authors found 14 associations of which 10 are extragalactic sources and 4 are of Galactic nature. Their main conclusion is that despite a large number of sources in the two catalogs, the very few associations are due to different emission mechanism in the two energy bands. However, even though the emission mechanism in both energy bands might not be the same, still the same source can be responsible for the emission in the two energy bands. Therefore, to successfully interpret the gamma-ray to hard X-ray association process a sizable number of sources is needed. This brings everything down to the sensitivity of both surveys. The more sensitive both surveys are, the more efficiently sources can be associated.\\
To improve the sensitivity of current hard X-ray surveys, we have applied the SIX survey to the entire sky
merging 60 months of BAT and 9 years of IBIS data. The resulting sources can be associated to the 2FGL sources matching a variety of different source classes. The approach is a nested source location constraining: the largest error radius of the gamma-ray source (from the LAT) is down-sized to 6 arcmin (by the SIX \cite{bottacini12}) that is further reduced to a few arcsec (by soft X-ray focusing telescopes). Thereby we are able to bridge the non-thermal (gamma-rays) and thermal (soft X-rays) universe making the association process physically justified. The hard X-ray to soft X-ray source association is a well established approach largely used in hard X-ray surveys e.g. \cite{bottacini12}.
Performances of the SIX survey show a maximum error radius in the source location of 6 arcmin and an angular resolution of 16 arcmin \cite{bottacini12}. \\
In order to find spatial coincidences we cross-correlate the SIX sources with the sky positions of the 2FGL sources over the entire sky. To account for the positional uncertainty of the {\em Fermi}-LAT sources, we find that the sources within $2\sigma$ confidence level have an uncertainty of less than 0.4 deg. The same approach was used by \cite{nolan12} for correlating 2FGL sources with BAT and IBIS separately. As a result with our approach with the 2FGL we obtain 84 associations. To test the efficiency of our approach, we have performed the same association but using randomized coordinates of the SIX sources. Only one correlation was found. This shows that the SIX-LAT sources do not correlate by chance and it translates into a mere 1.1\% of spurious associations. Thus, this result points to the fact that the association between hard X-rays and gamma-rays is physical. Out of our 84 associations 57 are extragalactic. In Figure~\ref{fig:extragal} we report the
flux -- flux plot of the extragalactic associated sources: X-axis is SIX flux, Y-axis is LAT flux. Color coding is: light green are blazars, dark green are TeV blazars, purple are non-blazars, red are previously unassociated sources.
% ------------- START f1.eps and f2.eps side by side---------------------------------------------------------
\begin{figure}[h]
\begin{minipage}{18pc}
\includegraphics[width=18pc]{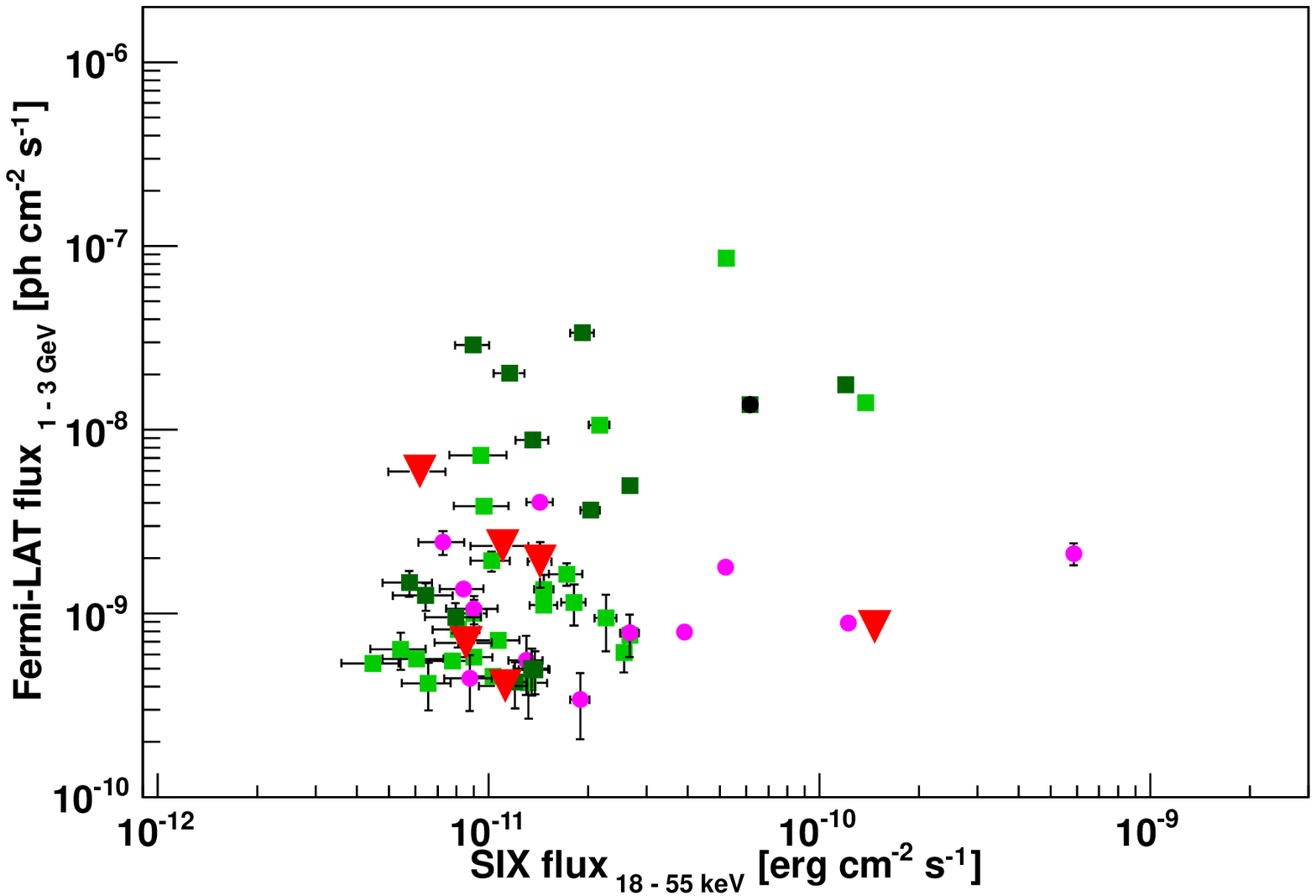}
\caption{\label{fig:extragal}SIX-flux vs LAT-flux of extragalactic associated sources. Color coding: light green are blazars, dark green are TeV blazars, purple are non-blazars, red are previously unassociated sources.}
\end{minipage}\hspace{1pc}%
\begin{minipage}{18pc}
\includegraphics[width=18pc]{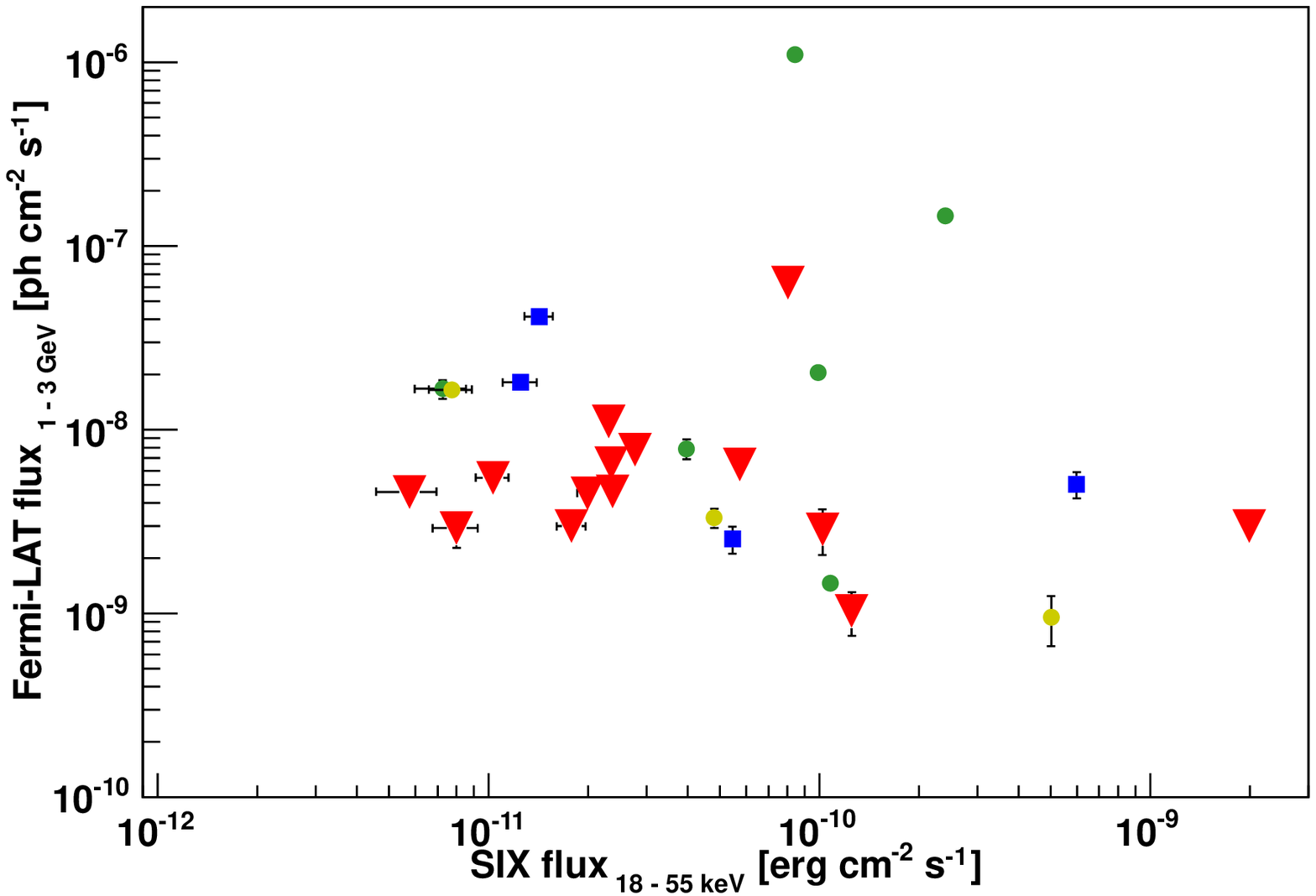}
\caption{\label{fig:gal}SIX-flux vs LAT-flux of galactic associated sources. Color coding: blue are HMXBs, green are pulsars, yellow are other type of sources, red are previously un-associated  sources.}
\end{minipage} 
\end{figure}
% ------------- END f1.eps and f2.eps side by side---------------------------------------------------------
%---------------------------- START spectrum 
\begin{figure}[ht]
\centering
\includegraphics[width=1.0\textwidth]{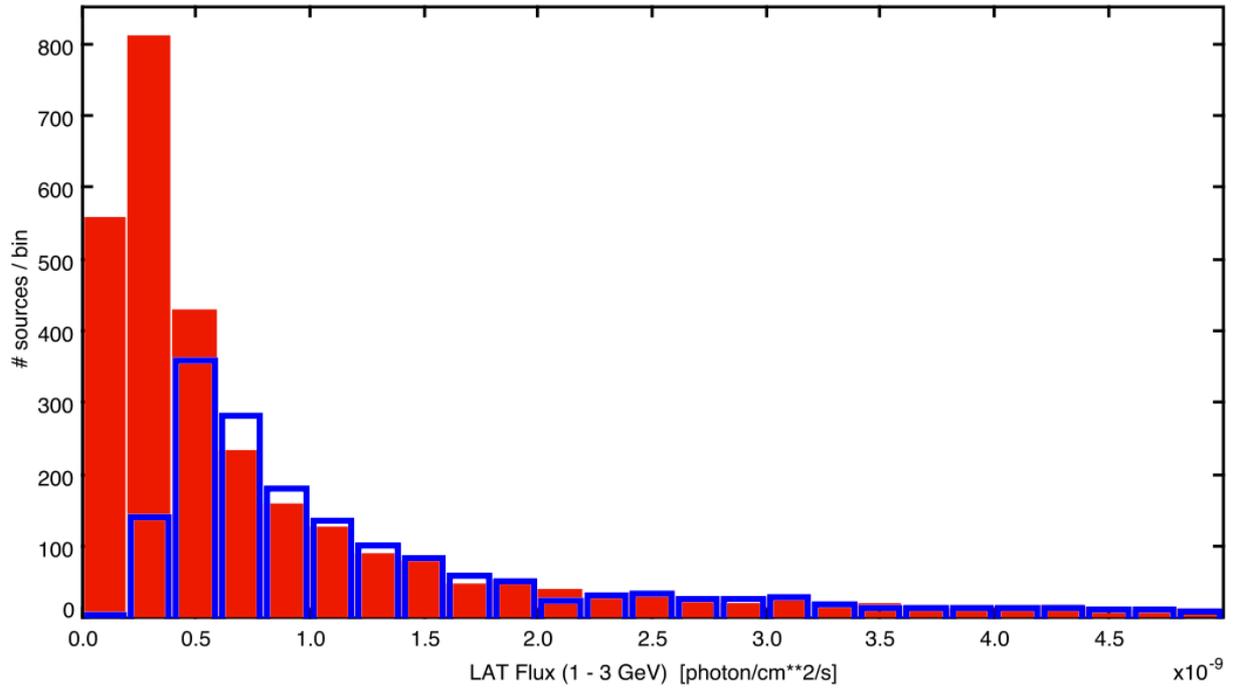}
\caption{Histogram plot displaying the number of sources (y-axis) per flux bin (x-axis) in units of $10^{-9}$ $photons$  $cm^{-2}$ $ s^{-1}$ of the 2FGL (blue empty histogram) and 3FGL (red filled histogram).}
\label{fig:2to3fgl}
\end{figure}
%---------------------------- END spectrum
Notably non-blazars (Seyferts and Radio Galaxies) are very weak emitters in the LAT energy range, while at SIX energies they span a wide range of fluxes. This is because they are common sources in the SIX survey \cite{bottacini12} that is in contrast to the LAT survey where they represent rare and interesting objects \cite{nolan12}.\\
Out of our 84 associations 27 are galactic. The flux -- flux plot of those sources is shown in Figure~\ref{fig:gal}. The color coding: blue are HMXBs, green are pulsars, yellow are other type of sources, red are previously un-associated  sources. Among the latter class there are: 1 AXP, 6 LMXBs, 3 HMXBs, 1 molecular cloud, 1 pulsar, 1 PWN, 1 new hard X-ray source without  any  counterpart.\\ \\
The SIX survey does not have any spurious source \cite{bottacini12}! Therefore, the association with a new hard X-ray source is a discovery of a potential as-yet-unknown type of {\em Fermi}-LAT gamma-ray source. It is worth emphasizing that the number of galactic hard X-ray -- gamma-ray source associations with our method has doubled with respect to already associated sources in the same energy bands in the 2FGL: the previously unassociated sources (red triangles) are 14 out of 27. There  is  no  apparent  correlation between the different galactic and extragalactic source classes. Yet the reason for this might be the still relatively low number of associated sources between the two energy bands.\\ \\
As the source association strongly depends on the sensitivity of the surveys, we compare the sources from the survey
of the 2FGL to the source catalog derived from the {\em Fermi}-LAT 4 years survey (3FGL) \cite{acero15}. To do this
we show a histogram plot in Figure~\ref{fig:2to3fgl}: the number of sources are shown on the y-axis as function of flux bin
on the x-axis in the 1 -- 3 GeV energy range. More than 1500 sources from the 3FGL (red filled histogram) are at
a flux level below 5$\times 10^{-10}$ $photons$ $cm^{-2}$ $s^{-1}$ where only ~300 sources in the 2FGL (blue empty histogram) are reported. This
improvement at low fluxes obtained with the 3FGL is very promising as most of the gamma-ray -- hard X-ray associations
occur at low fluxes.

\vspace*{0.5cm} 
\footnotesize{{\bf Acknowledgment:}{
E.B. acknowledges NASA grants NNX13AO84G and NNX13AF13G.}}

\end{document}